\documentclass[preprint2]{aastex}

\slugcomment{Submitted to the Astronomical Journal}

\shortauthors{Thorstensen, Fesen, and van den Bergh }
\shorttitle{Expansion Center and Age of Cas A}
\begin{document}

\title{The Expansion Center and Dynamical Age of the \\
        Galactic Supernova Remnant Cassiopeia A\thanks{Based in part
on observations with the NASA/ESA Hubble Space Telescope,
obtained at the Space Telescope Science Institute,
which is operated by the Association of Universities for Research in Astronomy,
Inc.
under NASA contract No. NAS5-26555.} }

\author{John R. Thorstensen \& Robert A. Fesen}
\affil{ 6127 Wilder Laboratory \\
Department of Physics \& Astronomy \\
Dartmouth College, Hanover, NH 03755}
\and
\author{Sidney van den Bergh}
\affil{Dominion Astrophysical Observatory \\
       Herzberg Institute of Astrophysics, National Research Council of Canada\\
       5071 West Saanich Road, Victoria, BC V9E 2E7, Canada}

\begin{abstract}

We present proper motions for 21 bright main shell and 17 faint, 
higher-velocity,
outer ejecta knots in the Cas A supernova remnant and
use them to derive new estimates for the remnant's expansion center and age.
Our study included $1951 - 1976$ Palomar 5~m prime focus plates,
$1988 - 1999$ CCD images from the KPNO 4~m and MDM 2.4~m telescopes,
and 1999 {\it{HST}} WFPC2 images. Measurable positions covered a 23 to 41 yr
time span for most knots, with a few outer knots followed for almost 48 yr.
We derive an expansion center of 
$\alpha$(J2000) = 23$^{\rm h}$ 23$^{\rm m}$ $27 \fs 77$ $\pm 0 \fs 05$,
$\delta$(J2000) = 58$^{\rm o}$ $48'$ $49 \farcs 4$ $\pm 0 \farcs 4$ (ICRS),
with little difference between centers derived using outer or main shell knots.
This position is $3 \farcs 0$ due north of that estimated
by \citet{vdBK83}.  It also lies
$6 \farcs 6$  $\pm 1 \farcs 5$  almost due north (PA = 354$^{\circ}$)
of the remnant's recently-detected central X-ray point source, implying
a transverse velocity for the X-ray point source $\simeq$ 330 km s$^{-1}$
at a distance of 3.4 kpc. Using the knots which lie out ahead of the
remnant's forward blast wave, we estimate a knot convergent date of A.D. 1671.3
$\pm$0.9 assuming no deceleration.  However, a deceleration of 
just $\sim 1.6$ km s$^{-1}$ yr$^{-1}$ over a 300 yr time span
would produce an explosion date $\simeq$ A.D. 1680, consistent with the 
suspected 
sighting of the Cas A supernova by J. Flamsteed.

\end{abstract}

\section{Introduction}

Cassiopeia A (Cas A) is the youngest Galactic supernova
remnant (SNR) known and, with the exception of the Sun, ranks as
the strongest discrete radio source in the sky at
$100 - 1000$ MHz.
At radio, optical, and X-ray wavelengths, Cas~A consists of a
$\simeq$ 2$'$ radius broken shell of SN debris rich in O, S, Si, Ar, Ca
expanding at 4000 -- 6000 km s$^{-1}$.
Within this shell lie about two dozen knots of much slower moving,
N-rich clumps of pre-SN, circumstellar mass loss material.
Outside of the shell, faint radio and X-ray emission extends
to a radial distance of $\simeq$ 160$''$ where a faint, filamentary edge
of X-ray emission marks the current location of the remnant's forward shock
front.
At an estimated distance of 3.4$^{+0.3}_{-0.1}$ kpc \citep{Reed95},
these angular dimensions correspond to main shell and outer
shock front radii of 2 pc and 2.7 pc respectively.
Several dozen faint optical knots with velocities of 8000 to 15,000 km s$^{-1}$
have been detected outside some sections of the main shell,
mainly in a northeastern ``jet'' of high speed ejecta
\citep{Fes96, Fes01}.

The remnant's precise age is uncertain.
From proper motion studies for $\sim$ 100 of Cas~A's optical knots
during 1951 -- 1980, \citet[hereafter KvdB76]{KvdB76}
and \citet[hereafter vdBK83]{vdBK83}
determined an explosion date of 1658 $\pm3$ for the
remnant as a whole (assuming no deceleration) and a somewhat later
date of 1671 for a few higher-velocity northeastern ``jet'' knots.
The difference between these derived dates probably reflects 
a greater deceleration of bright main shell knots
caused by their interaction with the remnant's reverse shock.  

There are no unambiguous historical
observations of a bright nova or variable star
in Cassiopeia that might be associated with a
late 17-th century supernova.
However, on 1680 August 16 John Flamsteed,
the first Astronomer Royal, reported seeing
a 5th -- 6th magnitude star he designated ``supra $\tau$'' and later
renamed 3 Cassiopeiae in his 1725 {\it Historia coelestia}
star catalog \citep{Ashworth80}.
Its proximity to Cas A, together with the fact that
he never observed this star again, raises the possibility that
he sighted the Cas A supernova in the summer of
1680.

The positional differences
between Cas A and the 3 Cas position are, however, troublingly large.
Flamsteed's location for 3 Cas is
offset from Cas A by $12 \farcm1$ in right ascension and $8 \farcm 6$
in declination.
Although refraction and sextant corrections might decrease these residuals
to $\simeq$ 6$'$ in both coordinates (errors not unprecedented
for Flamsteed), the case for Flamsteed's sighting of Cas A
is controversial \citep{Broughton79,Kamper80,Hughes80}.

Without additional evidence, the significance of Flamsteed's observation
might well remain inconclusive.
However, the large proper motions of the remnant's ejecta knots
($\mu$ = $0 \farcs 4 - 0 \farcs 6$ yr$^{-1}$) can be used to set
limits on Cas~A's age and thereby test the possible 1680 explosion date.
An accurate measurement of Cas A's age would in turn provide
key information about deceleration of its high-speed knots
and thus the phase of its evolutionary development.

Accurate proper motion measurements can also be used to improve
determinations of the the remnant's center of expansion.
KvdB76 determined the remnant's expansion center to
within an error radius of about one arcsecond.
No optical point source or extended emission
is present at this location down to I $\sim$ 24 mag \citep{vdBP86}.

Knowledge of Cas A's precise expansion center has recently
gained greater interest with the 
{\it Chandra X-Ray Observatory} discovery of an
X-ray point source near the remnant's center \citep{Tananbaum99}.
This object, which could be either a neutron star
with magnetized polar caps or an accreting compact
object \citep{Pavlov00,Umeda00,Chak01}, lies significantly offset from
estimates for the remnant's center of expansion (COE).
Point source coordinates derived from {\it ROSAT} and {\it Chandra} data
show a separation of $1'' - 5''$ from the KvdB76 and vdBK83 COE, and some
$16'' - 20''$ from the COE inferred by \citet{Reed95} using knot radial 
velocities.
These offsets imply transverse velocities of $50 - 250$ km s$^{-1}$
and $800 - 1000$ km s$^{-1}$ respectively, assuming d = 3.4 kpc and an age of
320 yr \citep{Pavlov00}.

In this paper, we present proper motions of 17
outlying high-velocity ejecta knots discovered over the last decade
[see \citet{Fes01} and references therein] along with 21
selected main shell knots. Many of these 40 knots can be seen on the earliest
archival Palomar 5~m PF plates, giving proper motion baselines
of nearly five decades, or about 1/7th of the remnant's age.
We use these proper motions
to determine a more accurate position and date for the supernova.

\section{Ejecta Knot Observations}

\subsection{Image Data}

Our observational material includes Palomar 5~m
prime focus (PF) plates dating back to 1951,
CCD images taken $1988 - 1999$ with the
KPNO 4~m and MDM 2.4~m telescopes, and a few
1999 {\it{HST} } WFPC2 images. Table 1 lists information
on the images that were used.
Although archival plates and modern CCD image data
have, in many cases, substantially different spectral sensitivities,
the angular scale of knot emission stratification or ionization structures 
lies well below all but the highest resolution {\it HST} data 
\citep{Fes02} and thus does not pose a significant problem for the 
inter-comparison of knot positions from these different data sets. 

To maximize the time base,
we examined several dozen archival Palomar 5~m PF
plates beginning with R. Minkowski in 1951
and ending with S. van den Bergh in 1989.
Most of these plates were unsuitable for this project
because of poor image quality or weak knot detection,
but four were selected for use.
These included two plates taken on back-to-back nights in 1951,
a better and deeper 1958 image (\citealt{vdBD70}, their Fig. 1),
and a superb 1976 image \citep{vdBK83}.

Modern CCD interference and broadband filter images
of all or portions of the Cas A
remnant obtained from November 1988 through October 1999 were also measured.
Several of these have been used in prior studies,
and a few were taken expressly for this project.
Some high resolution 1999 epoch WFPC2 {\it{HST} } images
were also used for several outlying northern and jet knots.

We selected for measurement a total of 38 ejecta knots
that had measurable positions
on the Palomar 1976 and later images ($\Delta$t = 23 yr).
Many were outer knots, together with some shell knots
visible from 1958 through 1999 ($\Delta$t = 41 yr).
Nearly a dozen knots, mostly among the outer ejecta, were detectable
from 1951 through 1999, a span of almost 48 years, which
is about 15\% of Cas A's age.

It is unlikely that suitable earlier images exist.
The Palomar 5~m was completed in 1948, coincident with the
discovery of Cas A as a localized radio source \citep{RS48}.
Furthermore, the remnant has brightened significantly over the
last half-century \citep{vdBK85}, and the individual knots
often have finite visibility lifetimes (KvdB76).
These factors set a practical limit of $\sim 50$ yr for
the time span over which the proper motion can be studied at
this time.

\subsection{Knot Selection}

Of the 38 knots selected, 21 were main
shell features and 17 were outer, higher-velocity
knots in the NE jet or elsewhere.  Figure 1 shows the location of the
selected  knots.
Table 2 cross-lists our designations with those of earlier studies
where possible.

\begin{figure}[t!] 
\epsscale{1.00}
\caption{ [S~II] $\lambda\lambda$6716,6731 image of Cas~A
from 1992 July \citep{Fes96}. The lower panel shows all the
knots used in this study.  The upper panel is a magnified view of the main
shell region.  The positions marked
are derived from the knot trajectories adjusted to the epoch of the image.
Some of the knots are not visible.}
\end{figure}

Our two main knot selection criteria were:
(1) {\it Distinct appearance with an absence of
significant morphological changes.}
We looked for knots which were compact with
steady morphologies that allowed secure identification and centroiding.
None of the knots was perfect but some, like the bright outer Knot 15,
provided excellent positional measurements over the entire 48 yr time span
surveyed. Like most other outlying fast-moving
knots, it has a relatively stable morphology (see \citealt{Fes01}).
On the other hand, main shell knots can show substantial changes in appearance
on images separated by just 5 to 10 yr. Consequently, 
we included only the most distinct and persistent 
main-shell features.  The number of main-shell features we
used was therefore relatively small compared to prior 
studies (KvdB76 and vdBK83).
(2) {\it Time span of visibility.}
The longer a knot is measurable,
the greater weight it has for determining the remnant's expansion center.
We therefore biased our knot selection toward knots with
long visibility time spans ($\geq$ 20 yr).
This resulted in a much smaller knot sample than the 102 measured by KvdB76.
Their knots covered
time spans ranging from $3 - 24$ yr, with 46\% of their knots
visible for less than 15 yr.

\section{Astrometric Procedures and Measurements}

\subsection{Reference Star Grid}

We began by constructing a grid of reference stars.
First, centroids for several hundred unsaturated
stars were measured on the H$\alpha$ and [S~II] $\lambda\lambda$6716,6731
images taken with the MDM
1.3~m telescope in 1992 July,
using the IRAF incarnation of DAOPHOT \citep{daophot}.
We then cross-identified these stars with the USNO A2.0 catalogue
\citep{USNOA2.0} and derived a 6-constant linear plate
model, rejecting stars with large residuals.
Because the centroids in the CCD data
have much better internal precision than the USNO A2.0
coordinates, we transformed the CCD centroids to
right ascension and declination using the plate model,
and averaged the results
from the H$\alpha$ and [S~II] exposures.
This procedure yielded right ascensions and declinations approximately
on the International Coordinate Reference System (ICRS) of the
USNO A2.0, but with much higher internal precision.
The USNO A2.0 is based on the original Palomar Observatory
Sky Survey plates, which for this field are epoch 1954.6,
so the elimination of high-residual stars
removes stars with appreciable proper motions,
as well as blended images and other difficult cases.
For the final step, the refined celestial
coordinates were converted to tangent-plane coordinates,
using the KvdB76 center of expansion as the tangent point.
The final reference grid
consisted of 141 stars with red magnitudes
(from the USNO A2.0) from 15.2 to 19.1 covering an
$8' \times 8'$ field.

Because this reference star grid is fundamental to all our results,
we checked it for various sources of error:  (1)
errors in the grid's {\it coordinate zero point} which would affect
comparisons with other results, (2) {\it proper motion} of the grid stars,
and (3)  {\it radial distortions} in the grid which could
cause a systematic error in the age estimate.
Below, we consider each of these sources of error in turn.

\subsubsection{Zero Point for Reference Grid}

We performed two checks on the positional zero point of the reference
star grid.

First, we examined the twelve
reference stars (for epoch 1965) tabulated by KvdB76.
We measured these stars on all the images on which they appeared,
used our reference star net to derive positions and
proper motion as described below for the knots, and compared
the results to those tabulated in KvdB76.
Ten of the 12 KvdB76 reference stars had sufficient
observations in our data.  For these, our right
ascensions were on average 485 mas larger than theirs, with an
RMS scatter of 67 mas, while our declinations were 448 mas
smaller, with a scatter of 112 mas.


Second, during the preparation of this work,
the Tycho-2 astrometric catalogue became available \citep{Hog2000}.
Four Tycho-2 stars appear on enough images to derive
good positions for epoch J2000, and for three stars the time base
gives adequate proper motions.  Because the stars were highly
saturated in nearly all our pictures, we estimated the star' centers
and their uncertainties by eye.  Our derived epoch 2000 positions for
the four Tycho stars are, on average $121 \pm 80$ mas east and
$82 \pm 79$ mas south of the catalogue positions, well within
the accuracy to which the USNO A2.0 is expected to
align with the ICRS.

The Tycho and USNO catalogues are based on Hipparcos observations, which
should provide much more reliable all-sky positions than the catalogs
available to KvdB76. We conclude that our grid is in registration with the ICRS
to within $\pm 0 \farcs 2$ at worst, and that KvdB may have suffered a barely
significant zero-point error.

\subsubsection{Proper Motions of Grid Stars}

In deriving plate models,
we did not adjust the positions of our grid stars for proper motions.
We implicitly assumed that the proper motions of the grid stars
were small.  Our grid stars are on average only a little brighter
than the dozen $V \sim 19$ mag
reference stars used by KvdB76.  They remark that stars this faint should have
intrinsic proper motions from the solar motion and Galactic
rotation of $\sim 1$ mas yr$^{-1}$.  It is therefore likely that our grid
stars' motions are similarly small.  A systematic
offset $\mu_{\rm sys}$ in proper motions can seriously affect the
derived center for the SNR, since it displaces the center
by  $\mu_{\rm sys} \times \sim 300\ \rm{yr}$.

For the three Tycho-2 stars for which
proper motions could be derived on our reference grid,
the weighted averages of
$\mu_{\rm T2} - \mu_{\rm grid}$ were
$-3.5 \pm 2.5$ mas yr$^{-1}$ in right ascension and
$+0.8 \pm 2.5$ mas yr$^{-1}$ in declination. Therefore, there
was no evidence for a significant motion with respect to the
ICRS.

The comparison with the KvdB reference stars was a bit
more complex.  They explicitly derived proper motions for their
reference stars, and found a 9 mas yr$^{-1}$ mean proper
motion in declination.  They decided this was spurious and
later adjusted their derived Cas A center to account for the
drift in their reference grid.  Reducing 10 of their
reference stars with respect to our grid gives mean
differences $\mu_{\rm KV} - \mu_{\rm ref}$
of $-2.8 \pm 1.4$ mas yr$^{-1}$ in right ascension and
$+8.3 \pm 1.4$ mas yr$^{-1}$ in declination.   This is
just as expected on the basis of their remarks.
Indeed, our procedure is logically similar to theirs, but
less complex: we simply assumed the faint grid to be motionless
{\it ab initio}, while they derived proper motions and
corrected them at the end to achieve the same result.

Finally, we checked our grid stars individually against the
reference grid, deriving proper motions for each star based
on all the images on which it appeared.
Formally, this was flawed by the inclusion of the star
itself in the plate models, but with our sample of 140 stars,
the effect should be negligible.  Grid stars observed over the whole
range of dates had typical estimated proper motion
uncertainties of $\pm \sim 1.7$ mas yr$^{-1}$, and relatively few showed
significant proper motions.  In the fitting
procedure used for the images (described below), grid
stars with large residuals were iteratively clipped out,
so the few stars with significant proper motions did not
affect our results.


From all these tests, we conservatively estimate the grid to be
inertial to within $\sim 2$ mas yr$^{-1}$.


\subsubsection{Field Distortions}

The procedure used to set up the reference
grid is valid provided that distortions in the
field of the MDM 1.3 m telescope are insignificant.  This is likely to be the
case. \citet{Cudworth91} measured the field distortions
of several southern telescopes, including the CTIO
1.5 m which, like the MDM 1.3 m, is an f/7.5 Ritchey-Chretien
reflector.  If one equipped the CTIO 1.5 m with a CCD
having the same size
as that used in deriving our reference grid,
then their radial distortion term $a_9$
would contribute only 7 mas at the corners
of the field of view.

Nonetheless, we searched for field distortions in
several different ways.  (1) In fitting the USNO A2.0
stars, we did not see any trends in the residuals
from the 6-constant plate model.  The USNO A2.0 typically
has centroiding errors in the 250 -- 500 mas range. In
view of the number of stars used, this alone limits
systematic trends to $\le 200$ mas.
(2) We examined archival CCD images of the globular cluster M13
taken with the MDM 1.3 m and the same camera as the Cas A
images.  Dr. Kyle Cudworth kindly provided us with a list of
star positions in M13, which he estimated were accurate to
$\sim 20$ mas for relative positions.  A six-constant fit of
the CCD centroids to
those positions gave an RMS residual of 70 mas, again
showing no obvious systematic trends. When a more
elaborate model was used, the residuals were not improved
significantly.  The scatter is somewhat larger than expected, but
does not seem to indicate any field distortions.
(3) In 1999 October we obtained a set of short
$I$-band CCD exposures of Cas A with the
MDM 2.4~m telescope, covering an $8' \times 8'$ field.
A 6-constant fit to the 124 standard reference stars included
in this image gave an RMS residual of 60 mas, without any iteration;
iterative clipping of high residuals brought this down to 34
mas, with 110 stars remaining.  Again, a more elaborate fit did
not result in significant improvement.

Because the reference
grid is based on exposures taken with the MDM 1.3 m,
this last test simply compares the two telescopes.
However, in January 2000
we also performed an astrometric calibration of the 2.4~m by
obtaining two sets of short-exposure 2.4~m images of a portion
of the \cite{stone99} astrometric standard region E.
In one set of images,
we fitted 172 stars to a 6-constant plate model. This
gave a 153 mas RMS residual. Iterative elimination of the largest residuals
brought this down to 53 mas with 114 stars.
Similar results were found with the other set of images.
The residual maps of the two sets of images
did not show systematic distortions, but were
highly correlated with each other. This suggests that most
of the error arises from the catalogue positions, probably due
to the (necessary) inclusion of many stars near the faint
limit of the catalogue.

In summary, the tests we made did not show any geometric distortion in
our reference grid.  The results suggest that systematic
distortions are smaller than $\sim 60$ mas and that the
centering precision of the reference stars is conservatively
$\sim 50$ mas.

\subsection{Image Solutions}

We scanned the four Palomar PF plates (Table 1) on the Yale
Astronomy Department's PDS microdensitometer.  We used
a $13.3\ \mu{\rm m} \times 13.3\ \mu{\rm m}$ scanning aperture
and sampled every 12.656 $\mu$m in a $3300 \times 4100$ raster
centered on the remnant, the long dimension being east-west.
The plate scale was $11 \farcs 1$  mm$^{-1}$, yielding $0 \farcs 141$
pixel$^{-1}$.  We used SExtractor (Bertin \& Arnouts 1996)
to derive star centers on for the Palomar plates.

The Palomar 5~m prime-focus camera used a corrector
which produced substantial radial distortions. Similar
corrector distortions have been discussed by \citet{murray71} and
\citet{Cudworth91}. Plate coordinates of the distortion
center, $(x_0, y_0)$, enter the least-squares models
in a non-linear fashion, so following \cite{murray71} and
\cite{Cudworth91} we estimated $(x_0, y_0)$
in a separate step.
We first fit the reference
stars with a simple 6-constant plate model of the form
$$X_0 = a_0 + a_1 x + a_2 y,$$
where $X_0$ is the standard reference star coordinate,
and $x$ and $y$ are the coordinates on the Palomar plate.
This fit gave root-mean-square (rms) residuals of $\sim 240$ mas, with
maximum values $\sim 690$ mas.
The distortion center  $(x_0, y_0)$ could then easily be estimated
from the residual maps.  These were then used in
a 16-constant model of the form
$$X_0 = a_0 + a_1 x + a_2 y + a_3 x^2 + a_4 xy + a_5 y ^2 + a_6 x r^2 +
	a_7 x r^4,$$
where $r^2 = ((x - x_0)^2 + (y - y_0)^2)^{1/2}$.  This
model is similar to that used by \citet{Cudworth91},
but without magnitude terms.  As expected, these fits
were much better; after a few of the highest-residual stars
were rejected,
rms residuals ranged from 71 mas (P7252) to 133 mas (P553B).

Fitting the CCD images was more straightforward.  We
centroided the reference stars with DAOFIND and matched
them to their standard star $XY$ coordinates. For the
ground-based CCD images, there were always $> 20$ stars
matched.  Because distortions in the CCD images were
expected to be relatively small, we used the 6-constant
model described above, with any unruly stars omitted by
iterative clipping.

Reduction of the few HST images was not as simple.
First, the images from the four
CCDs of the WFPC2 were interpolated onto a single grid using
the {\it wmosaic} task in the IRAF STSDAS package,
which approximately  corrects for field distortions.
Unfortunately, the WFPC2 field of view is so small that
one of our fields had only six reference stars.
To increase the number of reference stars we measured some
fainter stars on the `wide [S~II]' 2.4~m
image from 1999 October, transformed these over to the standard
grid and used them in the fit to the HST data.
With these added stars, each HST field had at least 14 reference stars.
The rms scatter for the 6-constant models was 50 to 55 mas.
\citet{wfpc2} quote an rms scatter of 10 mas in
their fit to the WFPC2 field distortions. The larger
scatter found here probably arises from the errors in the
reference star positions.  In particular, there was no
pattern indicating that the zero-point offsets between the
CCDs were different from those assumed by {\it wmosaic}.

\subsection{Knot Measurements and Fits}

We tried several methods to measure positions of the selected 38 knots.
This was a somewhat complicated problem, since many of the
knots were resolved in our images. The dominant
source of centering uncertainty resulted from structure in the
knots, rather than photon or grain noise.  In the end,
we simply estimated the knot centers by eye
using a cursor on an image display. We allowed our judgment
to be informed by centroids from such tools as the
{\tt imexamine} task in IRAF and (for the photographic
scans) the centers from {\tt sextractor}.  The selected
knots often had `head-tail' structures, in which case we
estimated the center of the head. Some of the shell knots were
embedded in nebulosity. In such cases, we tried to center on
the brightest part of the knot.

Our uncertainty estimates were also subjective,
but we attempted to err on the conservative side.
When we later fit straight-line trajectories to the knots,
we found that the residuals were often smaller than one would expect
on the basis of our estimated uncertainties, demonstrating
that our estimates were indeed conservative.
For the photographic images, our estimated errors were
guided by lower bounds on the
error based on the noise in the plate fog, the central
brightness of the knot, and the knot's angular size.
These bounds were calibrated with a Monte Carlo simulation.

Once $xy$ coordinates were measured on all the images,
they were transformed to the tangent-plane coordinate system
using the solutions described earlier. The rms scatter
in the plate solution was added in quadrature to each the
knot's estimated uncertainty. If the rms scatter was
less than 40 mas, it was set to 40 mas to account for
systematics. Transformed knot positions were automatically
collated with the exposure epochs for the images, yielding
a time series of positions for each knot.

Finally, each knot's trajectory was fitted with a
straight line,
$$X(t) = X_0(t_0) + \mu_X (t - t_0),$$
where $t_0$ is the weighted mean epoch of observation,
and similarly for $Y$.  Uncertainties were propagated
into the coefficients in a standard manner.
Table 3 gives the results of this procedure.

\section{Results and Discussion}

\subsection{Cas A's Center of Expansion}

Figure 2a shows the trajectories of our selected knots extrapolated
back to A.D. 1600, somewhat before the estimated explosion date.
The dots are the individual knot measurements with
the width of each line indicating its statistical weight.

\begin{figure}[t!] 
\epsscale{1.}
\plotone{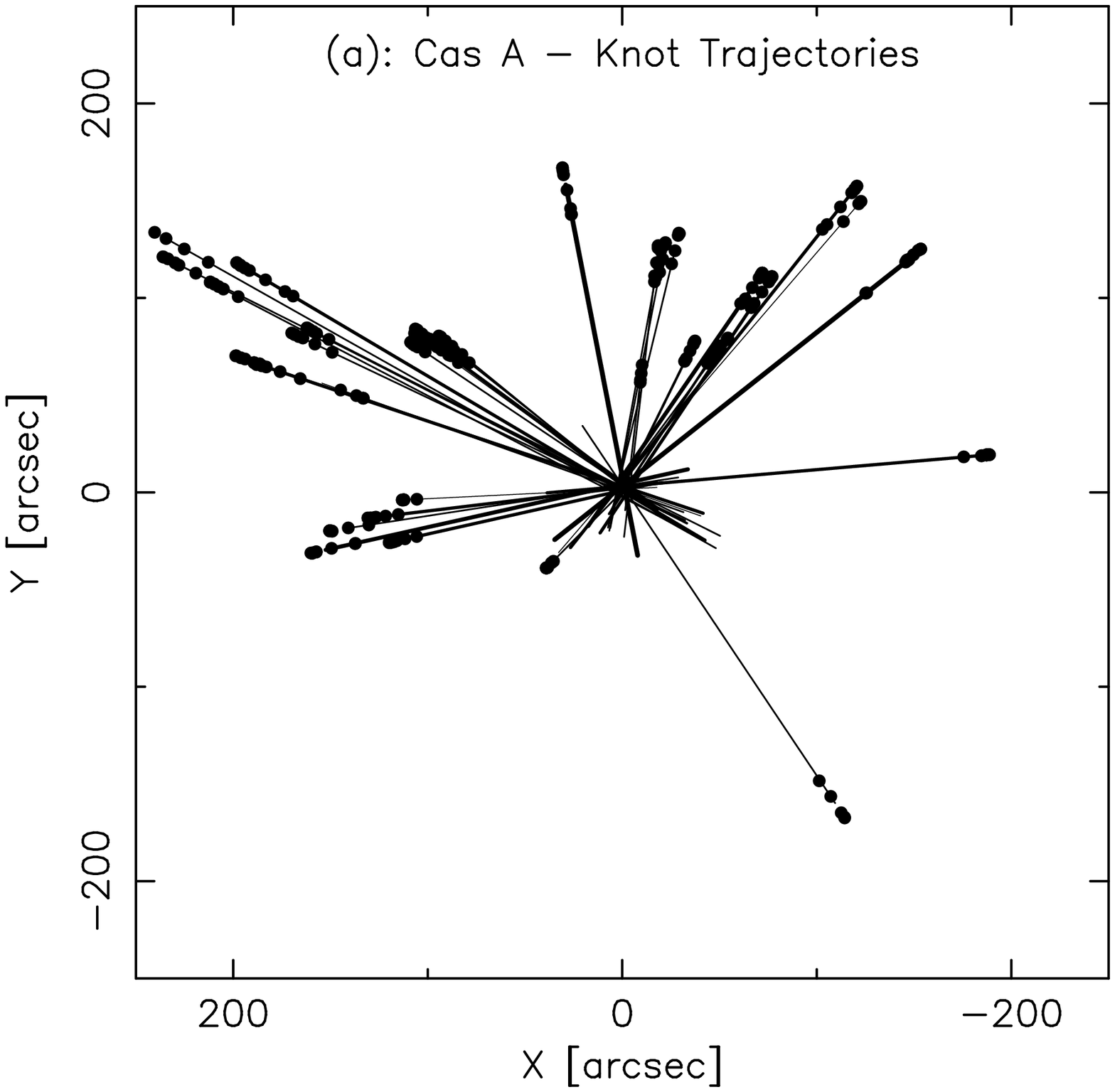}
\plotone{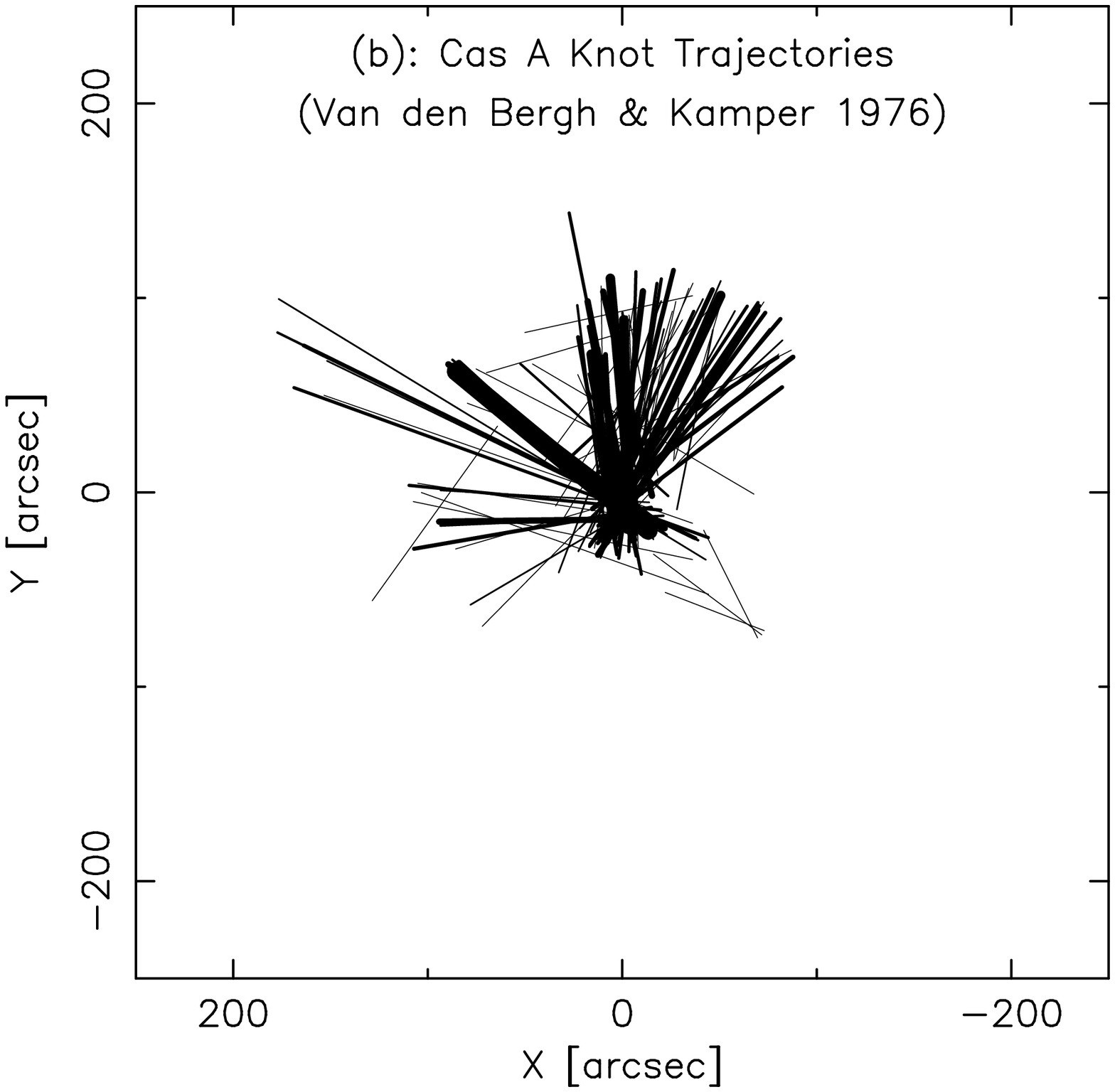}
\caption{a) Trajectories of the knots used in this study.
The dots are individual measurements, and the lines are fits to the
trajectories.
The widths of the lines increase according to the weight in the solution.
b) Trajectories of the 102 knots used by \citet{KvdB76}.}
\end{figure}

Because the knots may have decelerated by varying
amounts, we estimated the center using only the knots' lines
of position, and did not apply any constraints arising from
the time dependence (i.e, forcing the knots to start at the
same epoch).
We constructed a trajectory for each knot and computed
its positional uncertainty $\sigma_{i0}$ near the time of the explosion
by propagating the estimated position and proper motion errors.
Because of the long time lever, the proper motion uncertainty
dominated the errors in all cases.  With this information we could compute,
for any $X$ and $Y$, a likelihood function of the form
$$\lambda(X,Y) = \prod_i{{1 \over 2 \sigma_{i0}}
\exp(-d_{i \perp}^2 / 2 \sigma_{i0}^2)},$$
where $d_{i \perp}$ is the
perpendicular distance between $(X,Y)$ and the knot's line of
position.  The $(X,Y)$ which maximizes this is our estimate
of the expansion center.

This procedure gave $X = +91$ mas and
$Y = +2812$ mas, referred to the KvdB76 center. This translates into
$\alpha$(J2000) = 23$^{\rm h}$ 23$^{\rm m}$ $27 \fs 77$ $\pm 0 \fs 05$,
$\delta$(J2000) = 58$^{\rm o}$ $48'$ $49 \farcs 4$ $\pm 0 \farcs 4$.
The center derived using only the outer knots
nearly coincided with that derived from the selected shell knots.
Table 4 lists our main shell and outer knot centroids separately along
with our final values for the whole sample.
The errors given are purely statistical, based on the
Monte Carlo calibration of the centering errors (see below).

\subsubsection{Error Estimates}

Because our expansion center differs from those of previous estimates (see
below), we estimated our measurement uncertainties in several
ways.

Since we have computed a likelihood function,
the likelihood-ratio test
described by \citet{cash79} can be used to form confidence
contours. Such contours, however, assume that the positional
errors we estimated by eye are truly one standard deviation.
The resulting 95\%  confidence contour is an
oval slightly elongated northwest-southeast, with a
radius $\sim 1 \farcs 3$.  Figure 3 is a magnified view of the
center region, showing the lines of
position together with the 95\% to 99.5\% confidence contours
from this procedure.

\begin{figure}[t!] 
\epsscale{1.0}
\plotone{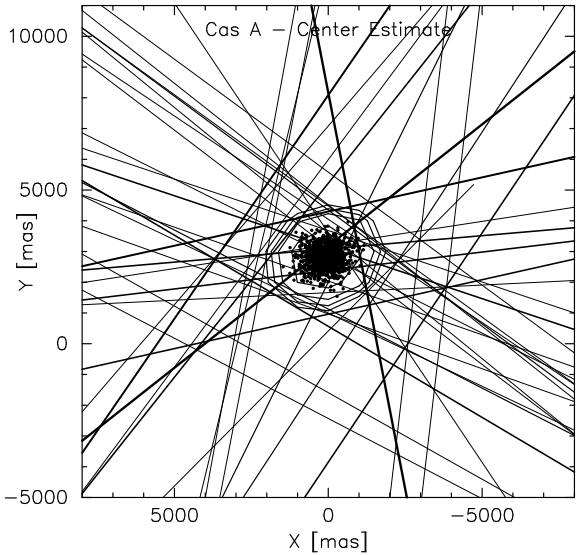}
\caption{ Magnified view of the region in which the knot trajectories
intersect. The straight lines are the fitted trajectories, again with the
line weight indicating the statistical weight.  The contours represent
confidence intervals for the centroid based on the likelihood ratio test, and
represent 95\% to 99.5\% confidence.  The dots are centroids determined from
Monte Carlo simulations of the measurement.  A `fudge factor' of 0.7 has been
applied to the estimated errors for this calculation (see text).  The origin
of this plot is the KvdB76 estimate of the center of expansion; note the
significant offset
of the new determination from this position. }
\end{figure}

We also used a Monte Carlo simulation of the proper motion measurements to
check our uncertainties and to normalize our error estimates.
We began by assuming that the explosion occurred
at the observed maximum-likelihood position.  Then, taking
each knot's present position as known and fixed (because
the proper motions dominate the errors),
we computed an idealized proper motion for each knot,
which extrapolated back exactly to the observed maximum-likelihood
position.  We next created 1000 artificial
data sets by adding Gaussian random noise to the idealized
proper motions.
For each artificial data set, we computed the maximum-likelihood center and
its associated maximum $\lambda$.

In the first trials,
the standard deviation used for the Gaussian noise
was simply the estimated proper motion uncertainty of each knot --
in effect, we took our estimated position errors to be realistic.
For nearly all the trials, this choice led to
maximum values of $\lambda$ larger than
observed, indicating that the errors in the proper motions were
overestimated (as expected since our error estimates were
believed conservative).  In the final Monte Carlo calculation,
we multiplied the proper motion standard
deviations by a global `fudge factor' $f \sim 0.7$.

These simulated data sets yielded maximum
values of $\lambda(X,Y)$ very similar to
that of the real data.  The spread in the positions
generated by this procedure should be a realistic indicator
of the uncertainty in the centroid.  The cloud of Monte Carlo
positions is also shown in Figure 3 and has standard deviations in
$X$ and $Y$ of 398 and 366 mas, respectively. Half of the points
lie within 448 mas of the mean position and 95\% within
946 mas.  This agrees well with the likelihood-ratio
test described above once the scaling factor of 0.7
is taken into account and provides a cross-check on the
likelihood-ratio procedure.

As an additional check, we divided our data into two samples --
shell knots and outer knots. The outer knots gave
$X = 599$ mas, $Y = 2809$ mas with half the Monte
Carlo points within 508 mas of the mean; the shell knots
gave $X = -388$ mas and $Y = 2840$ mas, with half within
783 mas.  The disagreements between these and the combined
data are about as expected given the estimated statistical errors.

In a final statistical experiment
we repeatedly selected half the knots at random found the
maximum likelihood center of the subsample.  For this
we used the observed trajectories without
adding artificial noise.
The distribution of the centers produced by 1000 such trials
was rather elongated east-west ($\sigma_X = 666$ mas, $\sigma_Y = 410$
mas), but did not extend toward a particular direction
(e.g., the KvdB76 and vdBK83 center). This shows
that our center determination is not thrown off
by a few stray knots.

\subsubsection{Comparison to Previous Results}

Previous center of expansion values from optical and radio
data are listed in Table 4.
The first accurate estimate of Cas A's expansion point was
made by \citet{vdBD70} using Palomar 5~m PF plates
of 27 ``fast-moving knots'' covering the time period 1951 through 1969.
\cite{KvdB76} later updated this to include a total of 102 knots
covering the additional period $1970 - 1975$.  This study itself was
supplemented by $1976 - 1980$ measurements of 46 especially long-lived knots by
\cite{vdBK83}. The last two studies reached essentially the
same central position estimate within measurement errors.
More recently, \cite{Reed95} found a significantly different expansion point
based on a least-squares spherical fit to a plot of main shell knot radial
velocities.
However, this displaced center reflects radial velocity differences
between back and front hemispheres and is thus unlikely to be an accurate 
measure of the remnant's expansion center.

Our estimated center lies only 3 arc seconds due north of the
center derived by KvdB76 and vdBK83, but this is well outside their
$\pm 0 \farcs 8 - 1 \farcs 0$ estimated uncertainty.
Although our study includes far fewer knots (38) than they used,
our results have smaller formal errors and the trajectories have a
smaller dispersion.
For comparison we show in Figure 2b trajectories
of the 102 knots measured in the KvdB76 study
at the same scale as the trajectories in Figure 2a.
The larger spread in knot trajectories in the KvdB76 data
largely reflects inclusion of fewer outlying knots
and more main shell knots in their study.

Finally, we note that the two reported centers derived from
radio measurements lie significantly ($\sim$15$''$) east
of the centers estimated from the optical knots.
Because the remnant's radio emission
does not exhibit a smooth, globally-coherent radial expansion
\citep{Bell77,Tuffs86,AR95},
we believe that radio-derived centers are less meaningful.

\subsubsection{X-ray Point Source}

First-light {\it Chandra} observations of Cas~A revealed the presence
of a point-like X-ray source near the center \citep{Tananbaum99}.
The source was subsequently confirmed through inspection of archival {\it ROSAT}
\citep{Aschenbach99} and {\it Einstein} data \citep{PZ99}.
The X-ray source's position has been refined slightly
using further {\it Chandra} observations \citep{murray01,
kap01}.
Table 5 lists this refined {\it Chandra} position, together
with those estimated from investigations of 
{\it ROSAT} and {\it Einstein} HRI data.
Figure 4 shows the {\it Chandra} position, and both KvdB76's and our 
expansion centers superposed on an optical
image of the central region.

\begin{figure}[t!] 
\epsscale{1.0}
\caption{Image of the center of the Cas A supernova
remnant taken with the MDM 2.4 m Hiltner telescope and a
[S~II] $\lambda\lambda$6716,6731 filter in $1''$ seeing.
The positions of the Chandra X-ray point source, the
KvdB76 expansion center, and our revised expansion center
are indicated.}
\end{figure}

The {\it Chandra} X-ray point source lies some $6 \farcs 6$ south
(position angle $\simeq$ 354$^{\circ}$) of
our derived center of expansion for Cas A.  Our new expansion center
actually lies {\it farther away} from the
{\it Chandra} position than does the KvdB76 center.
Both the {\it ROSAT} and
{\it Einstein} observations, however, place the X-ray point
source a few arc seconds
farther to the west and north, thus closer to our
expansion center.  Assuming a common origin for the
ejecta knots and the point source, our center and the
{\it Chandra} position together imply a transverse velocity
of $\simeq$ 330 km s$^{-1}$ at a distance of 3.4 kpc.  This velocity is not
unusual
for a young pulsar, if this is indeed what it is \citep{Umeda00,
Chak01,McL01}.

\subsection{Explosion Date and Knot Deceleration}

Limits on the date of the Cas A SN explosion bear on two questions.
First, could Flamsteed have seen it in A.D. 1680?
Second, how much have the knots decelerated over the last 300 yr?

The outer and shell knot trajectories give
significantly different ages, in the sense expected if the
shell knots have decelerated more than the outer knots due to 
their interaction with a reverse shock.
This trend is clearly visible in Figure 5, which shows the times at
which the individual knots are computed to have passed
closest to the center (assuming no deceleration);
we will refer to this quantity as the {\it crossing time}. (Note: The center
position used to compute the crossing times is derived from the
whole data set.  The knots move quickly enough that small
variations in the placement of the center do not affect the trend.)

\begin{figure}[t!] 
\epsscale{1.0}
\plotone{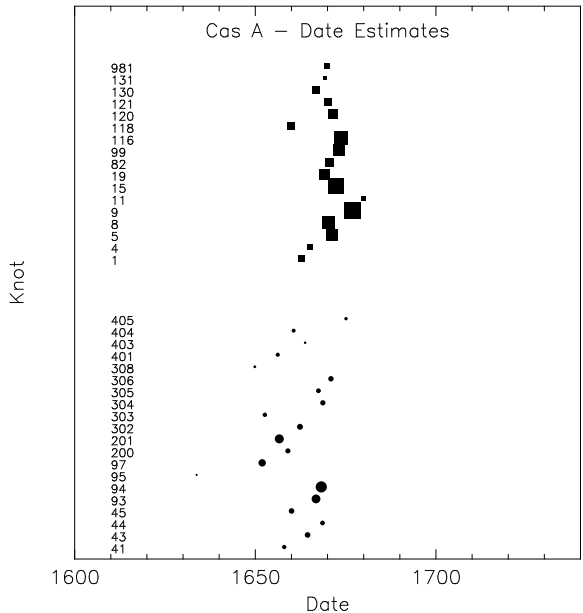}
\caption{ Dates at which the extrapolated knot trajectories pass nearest
to the maximum-likelihood center.  The size of the symbol is inversely
proportional
to the estimated uncertainty.  Outer knots (squares) are toward the top, and
shell knots (circles) toward the bottom.  Note the retardation and increased
scatter of the shell. }
\end{figure}

Because of their smaller deceleration, 
the outer knots offer a better estimate of the
date of the Cas A supernova than ejecta in the bright shell.  
A straight average of the 17 outer knots' crossing times yields an explosion 
date of
$1671.3 \pm 0.9$, while the 21 main shell knots yield $1662 \pm 1.7$.
On the face of it, the outer knot data
indicate a date nine years earlier than Flamsteed's
1680 sighting of 3 Cas.  However,
a deceleration of only $\sim 0.1$ mas yr$^{-2}$ for the
outer knots would change the date by $\sim 10$ years.
In our best-observed knots the 1$\sigma$ uncertainty
in the deceleration approaches this value, but
none of the knots show significant deceleration.
There is also no trend for the fitted acceleration
vectors to be pointed inward toward the center.
Thus we can neither directly detect nor disprove decelerations
large enough to make Flamsteed's A.D. 1680 sighting coincide with the
explosion.

The dispersion in the knots' crossing times also affects
the case for Flamsteed's 1680 sighting.  As noted above, the
mean crossing time is displaced from 1680.
If in addition the dispersion in crossing times were small,
then in order for Flamsteed to have seen the
explosion the different knots would need to have suffered
nearly identical decelerations.  This would be unlikely
given the typical inhomogeneity of the ISM.  However,
Figure 5 shows enough scatter in the crossing times
that a Flamsteed sighting remains plausible.

If we assume for a moment that the
explosion date really was 1680 and that the knots have
decelerated uniformly ever since, we can then
compute an implied deceleration for each knot.
For the nine best shell knots (those with
formal 1-$\sigma$ crossing time uncertainties less than
5 years), these implied decelerations range from
0.04 mas yr$^{-2}$ for Knot 9 to 0.14 mas yr$^{-2}$
for Knot 120, with a mean of 0.10 mas yr$^{-1}$.
At 3.4 kpc, 0.10 mas yr$^{-1}$ corresponds to a transverse
acceleration of only 1.6 km s$^{-1}$ yr$^{-1}$ or
a velocity change of some 2 -- 5\%
over the age of the remnant.

Detecting velocity changes $\sim 1 - 2$ km s$^{-1}$ yr$^{-1}$ in these faint, 
outer ejecta knots to test plausible explosion dates would be difficult, but 
perhaps not impossible. The sudden brightening of Knot 19 along the remnant's 
western limb during the early 1970's \citep{Fes01} suggests it may be 
decelerating significantly at the present epoch. 
A direct measurements of a knot's present-day deceleration could, 
in principle, be used to explore the density of the interstellar 
or circumstellar medium around the remnant.  But at present not 
enough is known about the knots' masses, dimensions, and structure to draw
reliable conclusions about their environment from their decelerations.

\subsection{Explicitly Time-Dependent Estimates}

As a check on the center and explosion date estimates,
we also estimated these quantities jointly.  Using the fitted knot
trajectories, we stepped through a range of dates $t$ around the explosion.
At each date, we computed the weighted mean position of the knots,
and then computed
$$S = \sum_i \left(d_i(t) \over \sigma_{i0}\right)^2,$$ 
where $d_i(t)$ is the angular distance of knot $i$ from the mean center
at date $t$.  The estimated explosion date is then $t_{\rm min}$,
the date which minimizes $S$.  The weighted mean position of
the knots at $t_{\rm min}$ is an estimate of the explosion center.
This procedure yielded $t = 1671.5$, $X = +723$ and $Y = +2362$ for the outer
knots, $t = 1659.8$, $X = +585$ and $Y = +1623$ for the shell knots, and 
$t = 1669.8$, $X = +856$, and $Y = +2693$ for the entire
sample of knots.  Because the knots in our sample are distributed 
non-uniformly around the remnant's periphery, differences in
date translate into differences in position in a complicated way.
The likely differential deceleration of the knots therefore makes this
estimate less reliable than the previous estimate, which is
based on lines of position.  Nonetheless, the results are
broadly similar -- all the estimates put the best center significantly
north of KvdB76's center, and the shell knots show evidence of
deceleration.

\section{Conclusions}

Our new proper motions of 21 main-shell and 17 higher-velocity,
outer ejecta knots in Cas A leads to improved
estimates of the center of expansion and the age.
We find the expansion center to be
$\alpha$(J2000) = 23$^{\rm h}$ 23$^{\rm m}$ $27 \fs 77$ $\pm 0 \fs 05$,
$\delta$(J2000) = 58$^{\rm o}$ $48'$ $49 \farcs 4$ $\pm 0 \farcs 4$,
with little difference between the centers derived using outer or main shell
knots. This expansion point lies $6 \farcs 6$  $\pm 1 \farcs 5$  to the
north of the recently recognized X-ray point source.  If the point source
originated in the explosion, the position offset implies
a transverse velocity of $\simeq$ 330 km s$^{-1}$
at a distance of 3.4 kpc.

Using the outer knots, most of which are in front of the
main blast wave, we estimate a date of explosion of 1671.3
$\pm$0.9
assuming no deceleration. However, interaction with local CSM/ISM
should decelerate the knots.  If the velocities have declined by only
a few percent over the age of the remnant, the remnant age would be
consistent with a suspected sighting of the
supernova by J. Flamsteed in 1680.
The age derived from the main shell knots is greater by
9 yr than that derived from the outer knots, implying
a greater deceleration of the main shell. 

\acknowledgments

We are grateful to R. Barr and the MDM staff for their
excellent assistance in instrument setup and preparation, to
the Dunlap Observatory at the University of Toronto for
shipment of several Palomar PF plates, and to C. Gerardy with
assistance with HST image data reduction.
Special thanks go to Terry Girard and John Lee at Yale for
their generous assistance with the plate scanning, and to the
Yale department for their hospitality.
One of us (SvdB) would like to thank the Mount Wilson
and Palomar Observatories for their kind
hospitality, and generous allocation of significant
amounts of 200-inch observing time over more than two
decades that were used to observe the motions in
Cassiopeia A. Van den Bergh would also like
to thank the late Karl Kamper for many
years of fruitful collaboration on the study of proper
motions and brightness variations of the knots in Cas A.
Partial support for this work was provided by NASA through grant number
GO-7405-01A
from the Space Telescope Science Institute,
which is operated by AURA, Inc., under NASA contract NAS5-26555.
JRT thanks the NSF for support through grant number 9987334.

\clearpage

\begin{deluxetable}{llccccl}
\tabletypesize{\scriptsize}
\tablecaption{Observational Material Used for Proper Motions }
\tablewidth{0pt}
\tablehead{
\colhead{Date} & \colhead{Telescope} & \colhead{Plate No./} &  \colhead{Scale}
&
\colhead{Emulsion/} &  \colhead{Region} &  \colhead{Exposures}  \\
\colhead{(U.T.)} & \colhead{} & \colhead{Image ID} &  \colhead{( $''$/pix)}  &
\colhead{Bandpass} & \colhead{Imaged}  & \colhead{(s)}
}
\startdata
1951 Nov 01 & Palomar 5 m &  553B  &0.1406 & 103aE + RG2      & Whole SNR  & 1 x
7200 \\
1951 Nov 02 & Palomar 5 m &  563B  &0.1406 & 103aE + RG2      & Whole SNR  & 1 x
7200 \\
1958 Aug 11 & Palomar 5 m & 3033S  &0.1406 & 103aF + RG2      & Whole SNR  & 1 x
5400 \\
1976 Jul 02 & Palomar 5 m & 7252vB &0.1406 & 098-04 + RG645   & Whole SNR  & 1 x
7200 \\
1988 Nov 10 & KPNO 4 m    & \nodata&0.297~ & [S II] $\lambda$6725& Jet, East
Limb & 5 x ~128 \\
1988 Nov 10 & KPNO 4 m    & \nodata&0.297~ & H$\alpha$+[N~II] & Jet, East Limb &
5 x ~192 \\
1992 Jul 05 & MDM 1.3 m   & \nodata&0.635~ & broad [S~II] $\lambda$6725& Whole
SNR & 3 x 1800 \\
1992 Jul 05 & MDM 1.3 m   & \nodata&0.635~ & H$\alpha$+[N~II] & Whole SNR & 3 x
1800 \\
1996 Oct 06 & MDM 2.4 m&\nodata&0.275~ &broad [S~II] $\lambda$6725& NW,SW,NE,SE
& 2 x 1000 \\
1996 Oct 07 & MDM 2.4 m&\nodata&0.275~ &H$\alpha$+[N~II]  &
NW,SW,NE,SE & 2 x ~600 \\
1996 Oct 07 & MDM 2.4 m&\nodata&0.275~ & Cont. 6450 \AA \ & NW,SW,NE & 2 x 1000
\\
1999 Jun 12 & HST 2.4 m&U52B0194R--0109R&0.0996 & [S~II] (F673N)     & Jet    &
6 x 1000 \\
1999 Jun 12 & HST 2.4 m&U52B010CR--010FR&0.0996 &
R (F675W)     & Jet    & 4 x ~600  \\
1999 Jun 13 & HST 2.4 m&U52B0205R--0207R&0.0996 &
R (F675W)     &NW Limb & 3 x ~500  \\
1999 Jun 13 & HST 2.4 m&U52B0205R--020BR&0.0996 &
R (F675W)     &West Limb & 2 x ~700 \\
1999  Oct 15 & MDM 2.4 m& \nodata& 0.275~        & R              &Whole SNR &
 2 x ~720   \\
\enddata
\tablenotetext{a}{Original Palomar 5 m plate scale: $11 \farcs 1$ per mm.}
\end{deluxetable}

\clearpage

\begin{deluxetable}{ccl}
\tabletypesize{\footnotesize}
\tablecaption{Proper Motion Knot Identification (ID) References}
\tablewidth{0pt}
\tablehead{
\colhead{Knot IDs }  & \colhead{Prior IDs for Knots} & \colhead{Reference}
}
\startdata
 \underline{Outer Knots}  &          &       \\
1,4,5,8,9     &1,4,5,8,9 & FBB87, FBG88 \\
   11         &   11     & FBG88 \\
   15         & KB91, 15 & vdBK83; FBB87; FBG88 \\
   19         &   19     & Fesen 2001 \\
 \underline{Knots in NE Jet}   &          &       \\
    82        &   82     & FB91 \\
    99        &  115     & vdBK83; FBG88 \\
   116        &  116     & KvB76, vdBK83; FBG88 \\
118,120,121   &118,120,121& FBG88 \\
   130        &  113     & vdBK83 \\
   131        &  ---     & this paper \\
\underline{Shell Knots}  &          &        \\
   93         &  41      & KvdB76, vdBK83 \\
   94         &   7      & KvdB76, vdBK83 \\
   95         &  ---     & this paper    \\
   97         &  111     & vdBK83 \\
  200         &  ---     & this paper \\
  201         &  ---     & this paper \\
  302         &  102     & KvdB76, vdBK83 \\
  303         &  ---     & this paper \\
  304         &  101     & KvdB76, vdBK83 \\
  305         &  105     & KvdB76, vdBK83 \\
  306         &  ---     & this paper \\
  308         &  ---     & this paper \\
  401         &  4       & vdBD70      \\
  403         &  65      & KvdB76 \\
  404         &  ---     & this paper \\
  405         &  ---     & this paper  \\
  41          &  59      &  KvdB76 \\
  43          &  14      &  KvdB76, vdBK83  \\
  44          &  ---     &  this paper \\
  45          &  63      &  KvdB76 \\
  501         &  1 ?     &  KvdB76 \\
  981         &  ---     &  this paper \\
\enddata
\tablenotetext{ ~}{References: FB91 = \citealt{Fes91};
                  FBG88 = \citealt{Fes88}; FBB87 = \citealt{Fes87};
                 KvdB76 = \citealt{KvdB76};
                 vdBD70 = \citealt{vdBD70};
                 vdBK83 = \citealt{vdBK83}.  }
\end{deluxetable}

\clearpage

\begin{deluxetable}{lccccccrr}
\tabletypesize{\scriptsize}
\tablecaption{Knot Positions and Proper Motions}
\tablewidth{0pt}
\tablehead{
\colhead{Knot} &
\colhead{$\alpha$} &
\colhead{$\delta$} &
\colhead{$\mu_\alpha$} &
\colhead{$\mu_\delta$} &
\colhead{$\sigma_x$} &
\colhead{$\sigma_\mu$} &
\colhead{First} &
\colhead{Last} \\
\colhead{ID} &
\colhead{(h\ \ m\ \ s)} &
\colhead{($^{\circ}\ \ '\ \ ''$)} &
\colhead{(mas yr$^{-1}$)} &
\colhead{(mas yr$^{-1}$)} &
\colhead{(mas)} &
\colhead{(mas yr$^{-1}$)} &
\colhead{Epoch} &
\colhead{Epoch}
}
\startdata
  1 & $ 23\ 23\ 12.943$ & $  58\ 45\ 57.97$ & $-341$ & $-509$ & $113$ & $ 10$ &
1958.61 & 1996.77 \\
  4 & $ 23\ 23\ 47.327$ & $  58\ 48\ 26.40$ & $ 454$ & $ -70$ & $172$ & $  9$ &
1951.83 & 1996.76 \\
  5 & $ 23\ 23\ 48.455$ & $  58\ 48\ 15.05$ & $ 488$ & $-108$ & $ 83$ & $  5$ &
1951.83 & 1999.79 \\
  8 & $ 23\ 23\ 12.202$ & $  58\ 51\ 24.06$ & $-370$ & $ 467$ & $ 74$ & $  5$ &
1951.83 & 1999.79 \\
  9 & $ 23\ 23\ 08.738$ & $  58\ 50\ 46.53$ & $-458$ & $ 362$ & $ 69$ & $  4$ &
1951.83 & 1999.79 \\
 11 & $ 23\ 23\ 11.921$ & $  58\ 51\ 16.43$ & $-391$ & $ 454$ & $143$ & $ 15$ &
1976.50 & 1999.45 \\
 15 & $ 23\ 23\ 31.727$ & $  58\ 51\ 33.51$ & $  97$ & $ 500$ & $ 77$ & $  4$ &
1951.83 & 1999.79 \\
 19 & $ 23\ 23\ 03.424$ & $  58\ 49\ 05.79$ & $-571$ & $  49$ & $ 58$ & $  6$ &
1976.50 & 1999.79 \\
 41 & $ 23\ 23\ 23.993$ & $  58\ 50\ 59.85$ & $ -92$ & $ 380$ & $181$ & $ 12$ &
1958.61 & 1999.79 \\
 43 & $ 23\ 23\ 25.187$ & $  58\ 50\ 52.39$ & $ -67$ & $ 365$ & $152$ & $  9$ &
1951.83 & 1999.79 \\
 44 & $ 23\ 23\ 25.367$ & $  58\ 50\ 53.47$ & $ -44$ & $ 376$ & $148$ & $ 10$ &
1958.61 & 1999.79 \\
 45 & $ 23\ 23\ 24.891$ & $  58\ 50\ 55.07$ & $ -71$ & $ 369$ & $150$ & $  9$ &
1958.61 & 1999.79 \\
 82 & $ 23\ 23\ 48.677$ & $  58\ 50\ 11.11$ & $ 491$ & $ 252$ & $ 72$ & $  8$ &
1976.50 & 1999.79 \\
 93 & $ 23\ 23\ 17.839$ & $  58\ 50\ 37.52$ & $-222$ & $ 331$ & $114$ & $  6$ &
1951.83 & 1999.79 \\
 94 & $ 23\ 23\ 18.480$ & $  58\ 50\ 39.43$ & $-225$ & $ 327$ & $ 97$ & $  5$ &
1951.83 & 1999.79 \\
 95 & $ 23\ 23\ 32.801$ & $  58\ 48\ 07.52$ & $ 110$ & $-111$ & $ 76$ & $ 10$ &
1958.61 & 1999.79 \\
 97 & $ 23\ 23\ 43.194$ & $  58\ 48\ 20.64$ & $ 345$ & $ -77$ & $115$ & $  6$ &
1958.61 & 1999.79 \\
 99 & $ 23\ 23\ 52.171$ & $  58\ 49\ 53.03$ & $ 580$ & $ 194$ & $ 48$ & $  6$ &
1958.61 & 1999.79 \\
116 & $ 23\ 23\ 53.316$ & $  58\ 50\ 44.52$ & $ 604$ & $ 358$ & $ 78$ & $  6$ &
1951.83 & 1999.79 \\
118 & $ 23\ 23\ 55.062$ & $  58\ 50\ 34.58$ & $ 616$ & $ 322$ & $ 89$ & $ 11$ &
1976.50 & 1999.79 \\
120 & $ 23\ 23\ 58.184$ & $  58\ 50\ 47.51$ & $ 719$ & $ 360$ & $107$ & $ 10$ &
1976.50 & 1999.79 \\
121 & $ 23\ 23\ 59.032$ & $  58\ 51\ 01.44$ & $ 730$ & $ 411$ & $107$ & $ 13$ &
1958.61 & 1996.76 \\
130 & $ 23\ 23\ 49.665$ & $  58\ 50\ 08.48$ & $ 508$ & $ 242$ & $141$ & $  9$ &
1958.61 & 1999.79 \\
131 & $ 23\ 23\ 53.600$ & $  58\ 49\ 57.28$ & $ 607$ & $ 203$ & $195$ & $ 21$ &
1976.50 & 1996.76 \\
200 & $ 23\ 23\ 41.778$ & $  58\ 50\ 03.71$ & $ 321$ & $ 215$ & $ 69$ & $ 10$ &
1976.50 & 1999.79 \\
201 & $ 23\ 23\ 44.598$ & $  58\ 48\ 33.20$ & $ 381$ & $ -46$ & $ 60$ & $  6$ &
1958.61 & 1999.79 \\
302 & $ 23\ 23\ 40.139$ & $  58\ 50\ 02.10$ & $ 287$ & $ 212$ & $126$ & $  8$ &
1958.61 & 1999.79 \\
303 & $ 23\ 23\ 39.275$ & $  58\ 50\ 01.73$ & $ 260$ & $ 205$ & $157$ & $ 10$ &
1958.61 & 1996.76 \\
304 & $ 23\ 23\ 39.899$ & $  58\ 50\ 07.15$ & $ 285$ & $ 233$ & $141$ & $  9$ &
1958.61 & 1999.79 \\
305 & $ 23\ 23\ 40.732$ & $  58\ 50\ 06.15$ & $ 307$ & $ 225$ & $162$ & $ 10$ &
1958.61 & 1999.79 \\
306 & $ 23\ 23\ 41.484$ & $  58\ 50\ 10.88$ & $ 324$ & $ 247$ & $143$ & $ 10$ &
1958.61 & 1999.79 \\
308 & $ 23\ 23\ 41.525$ & $  58\ 50\ 08.33$ & $ 307$ & $ 222$ & $170$ & $ 18$ &
1976.50 & 1999.79 \\
401 & $ 23\ 23\ 22.948$ & $  58\ 50\ 04.47$ & $-108$ & $ 219$ & $124$ & $  8$ &
1951.83 & 1999.79 \\
403 & $ 23\ 23\ 26.438$ & $  58\ 49\ 52.17$ & $ -22$ & $ 188$ & $140$ & $ 10$ &
1951.83 & 1999.79 \\
404 & $ 23\ 23\ 21.358$ & $  58\ 50\ 01.91$ & $-142$ & $ 217$ & $136$ & $  8$ &
1958.61 & 1999.79 \\
405 & $ 23\ 23\ 20.757$ & $  58\ 50\ 05.96$ & $-172$ & $ 232$ & $158$ & $ 11$ &
1958.61 & 1999.79 \\
501 & $ 23\ 23\ 42.343$ & $  58\ 48\ 42.61$ & $ 329$ & $ -16$ & $157$ & $ 14$ &
1976.50 & 1999.79 \\
981 & $ 23\ 23\ 47.840$ & $  58\ 49\ 43.04$ & $ 470$ & $ 168$ & $120$ & $ 11$ &
1951.83 & 1976.50 \\
\enddata
\tablecomments{Positions are for epoch J2000 and are referred to the ICRS. }
\end{deluxetable}

\clearpage

\begin{deluxetable}{lclll}
\tabletypesize{\scriptsize}
\tablecaption{Age and Center of Expansion Measurements for Cas A}
\tablewidth{0pt}
\tablehead{
\colhead{Reference} &
\colhead{SNR}    &
\colhead{T$_{\rm o}$ or}  &
\multicolumn{2}{c}{Expansion Center Coordinates} \\
\colhead{}          & \colhead{Region} & \colhead{SNR Age}
 &
          \colhead{$\alpha$(J2000)} &  \colhead{$\delta$(J2000)} }
\startdata
\underline{\bf{OPTICAL}}   &  &  &  & \\
\citealt{vdBD70}  &Brt. Shell & A.D. 1667 $\pm 8$ &
23$^{\rm h}$ 23$^{\rm m}$ $27 \fs 16$ $\pm 0 \fs 2$
&
 58$^{\rm o}$ $48'$ $47 \farcs 6$ $\pm 3 \farcs 1$ \\
\citealt{KvdB76} &Brt. Shell & A.D. 1653 $\pm 3$ &23$^{\rm h}$
23$^{\rm m}$ $27 \fs 76$ $\pm 0 \fs 1$
&
     58$^{\rm o}$ $48'$ $46 \farcs 7$ $\pm 0 \farcs 8$ \\
\ \ \ \ \ \ \ \ " ~ ~ ~ ~ ~ ~ ~ ~ "          & NE Jet    & A.D. 1671 $\pm 3$
&
              & \\
\citealt{vdBK83}  & Whole SNR& A.D. 1658 $\pm 3$ & 23$^{\rm h}$
23$^{\rm m}$ $27 \fs 76$ $\pm 0 \fs1$ &
     58$^{\rm o}$ $48'$ $46 \farcs 4$ $\pm 1 \farcs 0$ \\
\citealt{Fes88}     & Outer Knots&
     A.D. 1680 $\pm 15$ &  &  \\
\citealt{KvdB91} &Outer Knots&
     A.D. 1671 $\pm 3$ &  &  \\
\citealt{Reed95} &  Brt. Shell &
                   &
     23$^{\rm h}$ 23$^{\rm m}$ $26 \fs 55$ $\pm0\fs09$ &
     58$^{\rm o}$ $49'$ $00 \farcs 7$ $\pm0 \farcs 8$ \\
Thorstensen et al.  2001   & Outer Knots &
      A.D. 1671.3 $\pm$0.9   &
      23$^{\rm h}$ 23$^{\rm m}$ $27 \fs 84$ $\pm0\fs 06$ &
      58$^{\rm o}$ $48'$ $49 \farcs 4$ $\pm0 \farcs 4$ \\
   & Shell Knots &
      A.D. 1662.3 $\pm$1.7   &
      23$^{\rm h}$ 23$^{\rm m}$ $27 \fs 71$ $\pm0 \fs 09$ &
      58$^{\rm o}$ $48'$ $49 \farcs 5$ $\pm 0\farcs 7$ \\
   & Whole sample  &
      A.D. 1669.1 $\pm$0.8   &
      23$^{\rm h}$ 23$^{\rm m}$ $27 \fs 77$ $\pm0 \fs 05$ &
      58$^{\rm o}$ $48'$ $49 \farcs 4$ $\pm 0\farcs 4$ \\
\underline{\bf{RADIO}}                   &             &                   & &
  \\
\citealt{Tuffs86}         & Brt. Shell & Age = 949 yr  &
      23$^{\rm h}$ 23$^{\rm m}$ $25 \fs 95$ $\pm 0 \fs 4$ &
      58$^{\rm o}$ $48'$ $48 \farcs 4$ $\pm 4 \farcs 4$ \\
\citealt{Green88}   & Brt. Shell & Age $\sim$ 400 yr & &     \\
\citealt{AR95}      & Brt. Shell & Age $\sim$ 940 yr  &
      23$^{\rm h}$ 23$^{\rm m}$ $26 \fs 05$ $\pm 0 \fs 2$ &
      58$^{\rm o}$ $48'$ $54 \farcs 3$ $\pm 3 \farcs 1$ \\
\ \ \ \ \ \ \ \ \ " ~ ~ ~ ~ ~ ~  "          & Outer Knots &
      Age = $550 - 900$ yr & & \\
\citealt{AG99}  & Brt. Shell & Age = $400 - 500$ yr & &     \\

\underline{\bf{X-RAY}}    &            &                       & &     \\
\citealt{Vink98}  & Brt. Shell & Age = 501 $\pm$15 yr & &     \\
\citealt{Koralesky98}   & Brt. Shell & Age $\sim$ 500 yr     & &     \\
\enddata
\end{deluxetable}

\begin{deluxetable}{llll}
\tablecaption{Derived Center of Expansion versus X-ray Point Source }
\tablewidth{0pt}
\tablehead{
\colhead{Object} &
\colhead{Reference} &
\multicolumn{2}{c}{Coordinates} \\
\colhead{}     &   \colhead{}     & \colhead{$\alpha$(J2000)} &
\colhead{$\delta$(J2000)} }
\startdata
Center of Expansion & this paper &  23$^{\rm h}$ 23$^{\rm m}$ $27 \fs 77$ $\pm 0
\fs 05$ &
                                    58$^{\rm o}$ $48'$ $49 \farcs 4$ $\pm 0
\farcs 4$ \\
X-ray Pt. Source    & {\it Chandra} ACIS$^{a}$ &  23$^{\rm h}$ 23$^{\rm m}$ $27
\fs 86$ $\pm 0 \fs 13$  &
                                    58$^{\rm o}$ $48'$ $42 \farcs8 $ $\pm 1
\farcs 0$       \\
X-ray Pt. Source    & ROSAT HRI$^{b}$  & 23$^{\rm h}$ 23$^{\rm m}$ $27 \fs 57$
$\pm 0 \fs 75$  &
                                         58$^{\rm o}$ $48'$ $44 \farcs 0$  $\pm
6 \farcs 0$ \\
X-ray Pt. Source    & {\it Einstein} HRI$^{b}$  & 23$^{\rm h}$ 23$^{\rm m}$ $27
\fs 83$ $\pm 0 \fs 50$  &
                                          58$^{\rm o}$ $48'$ $43 \farcs 9$ $\pm
4 \farcs 0$      \\
X-ray Pt. Source    & {\it Einstein} HRI$^{b}$  & 23$^{\rm h}$ 23$^{\rm m}$ $27
\fs 89$ $\pm 0 \fs 50$  &
                                          58$^{\rm o}$ $48'$ $43 \farcs 7$ $\pm
4 \farcs 0$     \\
\enddata
\tablenotetext{a}{\cite{Tananbaum99}; \cite{kap01}; \cite{murray01}. }
\tablenotetext{b}{From \cite{Pavlov00}. }
\end{deluxetable}

\end{document}